# Angular momentum of light

A.M.Stewart

Research School of Physical Sciences and Engineering,
The Australian National University,
Canberra, Australia 0200. e-mail: andrew.stewart@anu.edu.au

**Abstract.** By means of the Helmholtz theorem on the decomposition of vector fields, the angular momentum of the classical electromagnetic field is decomposed, in a general and manifestly gauge invariant manner, into a spin component and an orbital component. The method is applied to linearly and circularly polarized plane waves in their classical and quantum forms.

## 1. Introduction

The notion of light containing orbital and spin components of angular momentum has become of interest in the past decade. For paraxial rays, that accurately describe laser light, the nature of the division into the two components appears now to be well understood [1-4].

However, for more general electromagnetic fields a simple, plausible and general derivation of these components from the expression for the angular momentum $\boldsymbol{J}(t)$ of the classical electromagnetic field in terms of the electric $\boldsymbol{E}(\boldsymbol{x},t)$ and magnetic field $\boldsymbol{B}(\boldsymbol{x},t)$ [5]

$$\boldsymbol{J}(t) = \frac{1}{4\pi c}\int \mathrm{d}^3x\, \boldsymbol{x} \times [\boldsymbol{E}(\boldsymbol{x},t) \times \boldsymbol{B}(\boldsymbol{x},t)] \tag{1}$$

(Gaussian units, bold font denotes a three-vector) seems to be lacking. Some authors [6] use decompositions that are not manifestly gauge invariant. Other decompositions lack generality [7]. In this paper, by applying the vector decomposition theorem of Helmholtz to the electric field, following [8], we obtain a decomposition of the angular momentum of the classical electromagnetic field into orbital and spin components that is quite general and manifestly gauge invariant throughout because it involves the fields only and not the potentials.

In section 2 we perform the Helmholtz decomposition, in section 3 we obtain general expressions for the orbital and spin parts of the classical electromagnetic field. In section 4 we apply the prescription to classical linearly and circularly polarized plane waves and discuss the apparent paradox that plane waves give rise to. In section 5 we show that when the prescription is applied to plane waves quantized in the Coulomb gauge it reproduces the standard forms for the orbital and spin angular momentum of the free photon. In section 6 we apply the same arguments to the linear momentum.

## 2. Helmholtz decompositions

The vector decomposition theorem of Helmholtz [9-11] states that any 3-vector field $\boldsymbol{E}(\boldsymbol{x})$ that vanishes at spatial infinity can be expressed, up to a uniform vector (see Appendix), as the sum of two terms

$$\boldsymbol{E}(\boldsymbol{x},t) = -\nabla_x f(\boldsymbol{x},t) + \nabla_x \times \boldsymbol{F}(\boldsymbol{x},t) \qquad (2)$$

where $\nabla_x$ is the gradient operator with respect to $\boldsymbol{x}$ and the scalar $f$ and vector $\boldsymbol{F}$ potential functions are

$$f(\boldsymbol{x},t) = \int \mathrm{d}^3y \, \frac{\nabla_y . \boldsymbol{E}(\boldsymbol{y},t)}{4\pi|\boldsymbol{x}-\boldsymbol{y}|} \quad \text{and} \quad \boldsymbol{F}(\boldsymbol{x},t) = \int \mathrm{d}^3y \, \frac{\nabla_y \times \boldsymbol{E}(\boldsymbol{y},t)}{4\pi|\boldsymbol{x}-\boldsymbol{y}|} \quad . \qquad (3)$$

The first term of (2) is known as the longitudinal part, the second as the transverse part. It is argued elsewhere [10-12] that the Helmholtz theorem applies to fields that vary with time. If the field is the electromagnetic $\boldsymbol{E}$ field then, after using Maxwell equations, the two potential functions become

$$f(\boldsymbol{x},t) = \int \mathrm{d}^3y \, \frac{\rho(\boldsymbol{y},t)}{|\boldsymbol{x}-\boldsymbol{y}|} \quad \text{and} \quad \boldsymbol{F}(\boldsymbol{x},t) = -\frac{\partial}{\partial t}\int \mathrm{d}^3y \, \frac{\boldsymbol{B}(\boldsymbol{y},t)}{4\pi c\,|\boldsymbol{x}-\boldsymbol{y}|} \qquad (4)$$

where $\rho(\boldsymbol{y},t)$ is the electric charge density and $\boldsymbol{B}$ the magnetic Maxwell field.

We see that the Helmholtz theorem decomposes the Maxwell electric field into gradients of the electromagnetic Coulomb gauge potentials $\boldsymbol{E} = - \quad f - \partial \boldsymbol{A}_t / \partial ct$. The expression for $\boldsymbol{A}_t$,

$$\boldsymbol{A}_t(\boldsymbol{x},t) = \quad {}_x \times \int d^3y \frac{\boldsymbol{B}(\boldsymbol{y},t)}{4\pi |\boldsymbol{x}-\boldsymbol{y}|} \tag{5}$$

was obtained previously [10, 11] by making a Helmholtz decomposition of the electromagnetic vector potential. The term $\boldsymbol{A}_t$ is the irreducible part of the electromagnetic vector potential that encodes all the information about the magnetic field [11]. The pure gauge term of the electromagnetic vector potential, which is the gradient of an arbitrary scalar field, does not encode any physical information.

When the electromagnetic ***E*** field is decomposed as in equation (2) one part of it, that involving the gradient in (2), is necessarily associated with the presence of electric charge (bound or b) and one part, that involving the curl in equation (2) (free or f), is not necessarily associated with the presence of electric charge. In this sense, the fields in an irregularly shaped metal cavity will be said to be *free* although they do not have the form of plane waves. The categorization applies to any physical quantities that depend on ***E***, such as the angular momentum.

The electromagnetic ***B*** field may also be decomposed by means of the Helmholtz theorem. Because $.\boldsymbol{B} = 0$, there is only one term in the decomposition and, with the use of the inhomogeneous Maxwell equation

$$\times \boldsymbol{B} = \frac{4\pi}{c}\boldsymbol{j} + \frac{1}{c}\frac{\partial \boldsymbol{E}}{\partial t} \tag{6}$$

where ***j*** is the electric current density, we get

$$\boldsymbol{B}(\boldsymbol{x},t) = -\frac{1}{c}\int d^3y\, \boldsymbol{j}(\boldsymbol{y},t) \times \quad {}_x \frac{1}{|\boldsymbol{x}-\boldsymbol{y}|} - \frac{1}{4\pi c}\int d^3y\, \frac{\partial \boldsymbol{E}(\boldsymbol{y},t)}{\partial t} \times \quad {}_x \frac{1}{|\boldsymbol{x}-\boldsymbol{y}|} \quad . \tag{7}$$

The first term is an instantaneous Biot-Savart term, the second term accounts for time dependence of the fields. These Helmholtz decompositions all have the feature that they are formally instantaneous in time.

## 3. Angular momentum of the classical electromagnetic field

To calculate the angular momentum of the classical electromagnetic field we decompose the ***E*** field according to (2) and (4) and substitute the result into (1).

### *3(a) Angular momentum of free fields*

We consider the term in (2) that contains the vector potential ***F***. With the use of a standard vector identity we expand the vector product as

$$( \ {}_x \times \boldsymbol{F}) \times \boldsymbol{B} = (\boldsymbol{B}.\ \ {}_x)\boldsymbol{F} - \sum_{r=1}^{3} B^r \ \ {}_x F^r \quad . \qquad (8)$$

The first term contributes to the angular momentum an amount

$$\boldsymbol{J}_{\mathrm{fs}} = \frac{1}{4\pi c}\int \mathrm{d}^3x\, \boldsymbol{x} \times (\boldsymbol{B}.\ \ {}_x)\boldsymbol{F} \qquad (9)$$

or, in components,

$$J_{\mathrm{fs}}{}^{i} = \frac{1}{4\pi c}\sum_{r,j,k=1}^{3} \varepsilon^{ijk}\int \mathrm{d}^3x\, x^j B^r \frac{\partial}{\partial x^r} F^k \qquad (10)$$

where $\varepsilon^{ijk}$ is the Levi-Civita tensor of rank 3. We do a partial integration with respect to $x^r$, assuming that boundary terms vanish, to get

$$J_{\mathrm{fs}}{}^{i} = -\frac{1}{4\pi c}\sum_{r,j,k=1}^{3} \varepsilon^{ijk}\int \mathrm{d}^3x\, F^k (\delta_{jr} B^r + x^j \frac{\partial}{\partial x^r} B^r) \qquad (11)$$

where $\delta$ is the Kronecker delta. The second term of (11) vanishes from **.*B*** = 0 and the first term gives

$$J_{\text{fs}}{}^{i} = -\frac{1}{4\pi c}\sum_{j,k=1}^{3}\varepsilon^{ijk}\int d^3x\, B^{j}F^{k} \tag{12}$$

which is the spin component of the angular momentum, in vector form

$$\boldsymbol{J}_{\text{fs}} = \frac{1}{4\pi c}\int d^3x\, \boldsymbol{F}\times\boldsymbol{B} \tag{13}$$

or explicitly

$$\boldsymbol{J}_{\text{fs}} = \frac{1}{(4\pi c)^2}\int d^3x\int d^3y\, \frac{\boldsymbol{B}(\boldsymbol{x},t)}{|\boldsymbol{x}-\boldsymbol{y}|}\times\frac{\partial \boldsymbol{B}(\boldsymbol{y},t)}{\partial t} \quad . \tag{14}$$

The second term of (8) cannot, by repeated partial integrations, be cast into a form that does not depend linearly on the vector ***x***. Accordingly, it gives the orbital component of the angular momentum of the free field

$$\boldsymbol{J}_{\text{fo}} = -\frac{1}{4\pi c}\int d^3x\, \boldsymbol{x}\times\sum_{r=1}^{3} B^{r} \quad {}_{x}F^{r} \quad . \tag{15}$$

By substituting for ***F*** from (4) and explicitly taking the gradient this may be expressed as

$$\boldsymbol{J}_{\text{fo}} = \frac{1}{(4\pi c)^2}\int d^3x\int d^3y\, [\boldsymbol{B}(\boldsymbol{x},t).\frac{\partial \boldsymbol{B}(\boldsymbol{y},t)}{\partial t}]\frac{\boldsymbol{x}\times\boldsymbol{y}}{|\boldsymbol{x}-\boldsymbol{y}|^3} \quad . \tag{16}$$

Equations (14) and (16) give, respectively, the spin and orbital components of the angular momentum of the free fields.

### *3(b) Angular momentum of bound fields*

The angular momentum of the bound fields, obtained from (1) and (2) is

$$\boldsymbol{J}_{\text{b}} = -\frac{1}{4\pi c}\int d^3x\, \boldsymbol{x}\times[\quad {}_{x}f\times\boldsymbol{B}] \quad . \tag{17}$$

By substituting for *f*, exchanging $\boldsymbol{x}$ and $\boldsymbol{y}$ and using the relation

$$\nabla_y(1/|\boldsymbol{y}-\boldsymbol{x}|) = -\nabla_x(1/|\boldsymbol{x}-\boldsymbol{y}|) \tag{18}$$

it becomes

$$\boldsymbol{J}_{\mathrm{b}} = \frac{1}{c}\int \mathrm{d}^3x\,\rho(\boldsymbol{x},t)\int\frac{\mathrm{d}^3y}{4\pi}\,\boldsymbol{y}\times\left[\nabla_x\times\frac{\boldsymbol{B}(\boldsymbol{y},t)}{|\boldsymbol{x}-\boldsymbol{y}|}\right] \quad . \tag{19}$$

The bound angular momentum has orbital character because it contains the coordinate vectors explicitly.

## 4. Plane waves: classical

It is desirable to verify that the equations developed in the last section are consistent with known results. To do this we apply them to the classical and quantum cases of plane waves of the electromagnetic field.

A linearly polarized classical plane wave propagating in the *z* direction has magnetic field $\boldsymbol{B}(\boldsymbol{x},t) = \hat{\boldsymbol{x}}B\sin(\omega t - kz)$ with time derivative $\partial\boldsymbol{B}(\boldsymbol{x}',t)/\partial t = \omega\hat{\boldsymbol{x}}B\cos(\omega t - kz')$ . There is also an electric field in the *y* direction. The cross product of the two vectors in (14) vanishes so the spin angular moment is zero. The scalar product is

$$\boldsymbol{B}(\boldsymbol{x},t).\partial\boldsymbol{B}(\boldsymbol{x}',t)/\partial t = \frac{\omega B^2}{2}\{\sin[k(z'-z)] + \sin[2\omega t - k(z+z')]\} \quad . \tag{20}$$

The time average of the second term vanishes and from (16) the time-averaged orbital angular momentum is

$$\boldsymbol{J}_{\mathrm{fo}} = \frac{1}{(4\pi c)^2}\int \mathrm{d}^3x \int \mathrm{d}^3x' \sin[k(z'-z)]\frac{\boldsymbol{x}\times\boldsymbol{x}'}{|\boldsymbol{x}-\boldsymbol{x}'|^3} \quad . \tag{21}$$

This integrand is odd under the transformation $\boldsymbol{x} \rightarrow -\boldsymbol{x}$ and $\boldsymbol{x}' \rightarrow -\boldsymbol{x}'$ and so the integral vanishes. Consequently both the spin and time-averaged orbital angular momentum of a linearly polarized plane wave vanish.

A circularly polarized plane wave propagating in the $z$ direction has magnetic field components

$$\boldsymbol{B}(\boldsymbol{x},t) = \hat{\boldsymbol{x}}B\cos(\omega t - kz) + \hat{\boldsymbol{y}}B\sin(\omega t - kz) \tag{22}$$

and from its time derivative we obtain from (14)

$$\boldsymbol{J}_{\mathrm{fs}} = \frac{\hat{\boldsymbol{z}}\omega B^2}{(4\pi c)^2}\int \mathrm{d}^3x \int \mathrm{d}^3x' \frac{\cos[k(z'-z)]}{|\boldsymbol{x}-\boldsymbol{x}'|} \quad . \tag{23}$$

The integral $\int \mathrm{d}^3x' \frac{\cos[k(z'-z)]}{|\boldsymbol{x}-\boldsymbol{x}'|}$ gives $4\pi/k^2$ so we get a vector density of spin angular momentum $B^2/4\pi\omega$ in the $z$ direction. In calculating the orbital angular momentum from (16) we get $\boldsymbol{B}(\boldsymbol{x},t).\partial \boldsymbol{B}(\boldsymbol{x}',t)/\partial t = \omega B^2\{\sin[k(z'-z)]\}$ and hence an expression similar to (21). The orbital angular momentum therefore vanishes. We find that the circularly polarized classical plane wave has a finite volume density of spin angular momentum $B^2/4\pi\omega$ but a zero volume density of orbital angular momentum. The energy density of the wave is $B^2/4\pi$ so the ratio of spin angular momentum density to energy density is $1/\omega$.

It is well known on the basis of experiment [13] and theory [14] and is confirmed by the calculation above that a circularly polarized plane wave carries spin angular momentum whereas a linearly polarized one does not. At first sight this seems inconsistent with (1) because, from the vector form of that equation, the angular momentum density can have no component in a direction perpendicular to the plane of $\boldsymbol{E}$ and $\boldsymbol{B}$. The matter has been the subject of discussion for a long time [15] and even recently [16-19]. The resolution of the paradox is that an obstacle that absorbs angular momentum changes the electromagnetic field at the edges of the obstacle so as to give the fields in that region a component in the direction of propagation. This gives rise to a Poynting vector that is not parallel to the direction of propagation and to an angular momentum vector that is. The issue has been discussed most clearly by Simmons and Guttman [20] who explain qualitatively that by partial integration the angular momentum-producing effects that occur at the edges of an obstacle may be viewed as occurring over the volume of the obstacle. It is this partial integration that has been carried out in a more complete way in section 3 of this paper.

An examination of the surface term that arises as a result of the partial integration of (10) shows [23] that for a circularly polarized wave it contributes a term that is equal and opposite to that given by the volume term, giving a total angular momentum of zero, in accordance with the predictions of (1). However, if the infinite wave is constricted by an aperture of finite size and the surface of integration is taken just outside the boundary of a section of the constricted wave then the surface term vanishes and only the volume term remains. In this way [24] a reconciliation is effected between the naïve predictions of (1) and the experiments of Beth [13].

## 5. Plane waves: quantum mechanical

A photon is the quantized normal mode of the electromagnetic field and, as such, the division of its angular momentum into spin and orbital parts depends on the shape of the particular normal mode. In order to compare our results with known ones we again consider the simplest case of plane waves, this time quantized in the Coulomb gauge (div $\boldsymbol{A}$ = 0, zero scalar potential). The vector potential of such a linearly polarized wave is

$$\boldsymbol{A}_{\mathrm{t}}(\boldsymbol{x},t)=\left(\frac{2\pi c\hbar}{V}\right)^{1/2}\sum_{\boldsymbol{k}}\frac{1}{\sqrt{|\boldsymbol{k}|}}\sum_{\lambda=1}^{2}\hat{\varepsilon}(\boldsymbol{k},\lambda)(a_{\boldsymbol{k},\lambda}e^{-ik.x}+a^{\dagger}_{\boldsymbol{k}\lambda}e^{ik.x}) \tag{24}$$

where $V$ is the normalization volume [21]. The polarization vectors are taken to satisfy the conditions $\hat{\varepsilon}(\boldsymbol{k},1)\times\hat{\varepsilon}(\boldsymbol{k},2)=\hat{\boldsymbol{k}}, \hat{\varepsilon}.\boldsymbol{k}=0,\ \hat{\varepsilon}(\boldsymbol{k},\lambda).\hat{\varepsilon}(\boldsymbol{k},\lambda')=\delta_{\lambda\lambda'},\ \hat{\varepsilon}(-\boldsymbol{k},1)=-\hat{\varepsilon}(\boldsymbol{k},1)$ and $\hat{\varepsilon}(-\boldsymbol{k},2)=\hat{\varepsilon}(\boldsymbol{k},2)$. The creation and destruction (Fock) operators obey the commutation relations $[a_{\boldsymbol{k}\lambda},a^{\dagger}_{\boldsymbol{k}',\lambda'}]=\delta_{\boldsymbol{k},\boldsymbol{k}'}\delta_{\lambda,\lambda'}$; other combinations commute. The Fock operators are normalized to give energy differences $\hbar\omega_{\boldsymbol{k},\lambda}$ for a given mode. The quantities $x = (ct, \boldsymbol{x})$ and $k = (\omega/c, \boldsymbol{k})$ are four-vectors so that their scalar product is $k.x=\omega t-\boldsymbol{k}.\boldsymbol{x}$.

*4(a) Second-quantized spin angular momentum of a plane wave photon*

The spin angular momentum is given by (14). From (24) we get

$$\boldsymbol{B}(\boldsymbol{x},t)=i\left(\frac{2\pi c\hbar}{V}\right)^{1/2}\sum_{\boldsymbol{k}}\frac{1}{\sqrt{|\boldsymbol{k}|}}\sum_{\lambda=1}^{2}\boldsymbol{k}\times\hat{\varepsilon}(\boldsymbol{k},\lambda)(a_{\boldsymbol{k},\lambda}e^{-ik.x}-a^{\dagger}_{\boldsymbol{k},\lambda}e^{ik.x}) \tag{25}$$

and

$$\frac{\partial\boldsymbol{B}(\boldsymbol{y},t)}{\partial t}=c\left(\frac{2\pi c\hbar}{V}\right)^{1/2}\sum_{\boldsymbol{k}'}\sqrt{|\boldsymbol{k}'|}\sum_{\lambda'=1}^{2}\boldsymbol{k}'\times\hat{\varepsilon}(\boldsymbol{k}',\lambda')(a_{\boldsymbol{k}',\lambda'}e^{-ik'.y}+a^{\dagger}_{\boldsymbol{k}',\lambda'}e^{ik'.y})\quad . \tag{26}$$

From substituting the product of (25) and (26) into (14) we find

$$\begin{aligned}\boldsymbol{J}_{\mathrm{fs}}=\left(\frac{i\hbar}{8\pi V}\right)\int d^3x\int\frac{d^3y}{|\boldsymbol{x}-\boldsymbol{y}|}\sum_{\boldsymbol{k},\boldsymbol{k}'}\sqrt{\frac{|\boldsymbol{k}'|}{|\boldsymbol{k}|}}\sum_{\lambda,\lambda'}\boldsymbol{k}'[\boldsymbol{k}.\hat{\varepsilon}(\boldsymbol{k},\lambda)\times\hat{\varepsilon}(\boldsymbol{k}',\lambda')]\\ \times(a_{\boldsymbol{k},\lambda}e^{-ik.x}-a^{\dagger}_{\boldsymbol{k},\lambda}e^{ik.x})(a_{\boldsymbol{k}',\lambda'}e^{-ik'.y}+a^{\dagger}_{\boldsymbol{k}',\lambda'}e^{ik'.y})\end{aligned}\quad . \tag{27}$$

For the terms containing $a^{\dagger}a$ and $aa^{\dagger}$, the integrals over the spatial coordinates $d^3x d^3y$ give a factor $4\pi V\delta_{\boldsymbol{k}',\boldsymbol{k}}/\boldsymbol{k}^2$. Since $\boldsymbol{k}' = \boldsymbol{k}$ it follows, from the presence of the vector cross product, that any non-zero terms in (27) must have $\lambda' \neq \lambda$. From the properties of the polarization vectors we find, for the linearly polarized photon, $\boldsymbol{J}_{\mathrm{fs}}=i\hbar\sum_{\boldsymbol{k}}\hat{k}(a^{\dagger}_{\boldsymbol{k},1}a_{\boldsymbol{k},2}-a^{\dagger}_{\boldsymbol{k},2}a_{\boldsymbol{k},1})$. This operator has zero expectation value in single photon states that are linearly polarized. However, by transforming to a circularly polarized basis in the usual way [22], we get the standard expression for the diagonal spin angular momentum operator of a circularly polarized photon

$$\boldsymbol{J}_{\mathrm{fs}} = \hbar \sum_{\boldsymbol{k}} \hat{\boldsymbol{k}} (a^{\dagger}_{\boldsymbol{k},\mathrm{R}} a_{\boldsymbol{k},\mathrm{R}} - a^{\dagger}_{\boldsymbol{k},\mathrm{L}} a_{\boldsymbol{k},\mathrm{L}}) \tag{28}$$

where R and L refer to photons that are right and left circularly polarized with respect to the wave vector $\boldsymbol{k}$.

For the terms with Fock operators aa and $a^{\dagger}a^{\dagger}$, the spatial integrals produce a factor $4\pi V \delta_{\boldsymbol{k}',-\boldsymbol{k}} / \boldsymbol{k}^2$. The condition $\boldsymbol{k}' = -\boldsymbol{k}$ also leads to $\lambda' \neq \lambda$ from the properties of the polarization vectors and, noting that the summands over $\boldsymbol{k}$ are odd in $\boldsymbol{k}$, these terms vanish. The only surviving term of the spin angular momentum operator is (28). It is therefore not necessary to normal-order the Fock operators to obtain (28).

*4(b) Second-quantized orbital angular momentum of a plane wave photon*

We integrate (15) by parts to get the $i$ th component of the free orbital angular momentum

$$\boldsymbol{J}^{i}_{\mathrm{of}} = -\frac{\varepsilon^{ijk}}{(4\pi c)^2} \sum_{r=1}^{3} \int d^3x \int d^3y \frac{x^j}{|\boldsymbol{x}-\boldsymbol{y}|} \frac{\partial B^r(\boldsymbol{x},t)}{\partial x^k} \frac{\partial B^r(\boldsymbol{y},t)}{\partial t} \tag{29}$$

and, by substituting from (25, 26),

$$\begin{aligned} \boldsymbol{J}^{i}_{of} = \frac{\varepsilon^{ijk}\hbar}{8\pi V} \sum_{\boldsymbol{k},\boldsymbol{k}',\lambda,\lambda'} \int d^3x \int d^3y \frac{k^k x^j}{|\boldsymbol{x}-\boldsymbol{y}|} \sqrt{\frac{k'}{k}} \frac{\boldsymbol{k}.\boldsymbol{k}'}{k'^2} \hat{\varepsilon}(\boldsymbol{k},\lambda).\hat{\varepsilon}(\boldsymbol{k}',\lambda')[a^{\dagger}_{\boldsymbol{k}\lambda} a_{\boldsymbol{k}',\lambda'} e^{ict(k-k')} e^{-i\boldsymbol{x}.(\boldsymbol{k}-\boldsymbol{k}')} \\ + a_{\boldsymbol{k}\lambda} a^{\dagger}_{\boldsymbol{k}',\lambda'} e^{-ict(k-k')} e^{i\boldsymbol{x}.(\boldsymbol{k}-\boldsymbol{k}')} + a_{\boldsymbol{k},\lambda} a_{\boldsymbol{k}',\lambda'} e^{-ict(k+k')} e^{i\boldsymbol{x}.(\boldsymbol{k}+\boldsymbol{k}')} + a^{\dagger}_{\boldsymbol{k},\lambda} a^{\dagger}_{\boldsymbol{k}',\lambda'} e^{ict(k+k')} e^{-i\boldsymbol{x}.(\boldsymbol{k}+\boldsymbol{k}')}]. \end{aligned} \tag{30}$$

After carrying out integrations of the form $\int \mathrm{d}^3 y \frac{e^{i\boldsymbol{k}'.(y-x)}}{|\boldsymbol{x}-\boldsymbol{y}|} = \frac{4\pi}{\boldsymbol{k}'^2}$, we end up with terms that have integrals like $\int_{-L/2}^{L/2} \mathrm{d}x\, x e^{iqx}$, where $L$ is the length of the side of the normalization cube. The real part of this integral is zero. The imaginary part oscillates rapidly as $L \to \infty$ and it is stationary only when $q \to 0$. However, in this limit, the integral itself vanishes and so does the orbital angular momentum. The reader may object that if the origin of coordinates is not chosen to be at the centre of the normalization volume a non-zero result will be obtained. However, if it is accepted that the vacuum is invariant under spatial inversion, then any simulation of the vacuum, such as by a normalization box, must be invariant under inversion too. Accordingly, the origin of coordinates must be taken to be at the centre of the box and the integral vanishes. We find that the second-quantized orbital angular momentum of a plane wave is zero, as expected from the classical situation discussed in the previous section. The correspondence with the classical situation is complete. The second-quantized angular momentum of bound fields is obtained by substituting (25) in (19).

## 6. Linear momentum

The components of the linear momentum of the electromagnetic field

$$\boldsymbol{P}(t) = \frac{1}{4\pi c}\int \mathrm{d}^3 x\, \boldsymbol{E}(\boldsymbol{x},t) \times \boldsymbol{B}(\boldsymbol{x},t) \tag{31}$$

may also be obtained from (2). By using (8) and, carrying out integrations by parts with vanishing surface terms, we get for the free component

$$\boldsymbol{P}_{\mathrm{f}}\,|^{i} = -\frac{1}{(4\pi c)^2}\int \mathrm{d}^3 x \int \mathrm{d}^3 y \sum_{r=1}^{3} \frac{1}{|\boldsymbol{x}-\boldsymbol{y}|} \frac{\partial \boldsymbol{B}^r(\boldsymbol{x},t)}{\partial x^i} \frac{\partial \boldsymbol{B}^r(\boldsymbol{y},t)}{\partial t} \quad . \tag{32}$$

For a circularly polarized plane wave the volume density of linear momentum given by (32) is found to be $\boldsymbol{B}^2/4\pi c$; for a linearly polarized plane wave the time average of the volume density is $\boldsymbol{B}^2/8\pi c$.

When (32) is quantized using (25, 26) it becomes, after a routine calculation, the standard expression

$$\boldsymbol{P}_{\mathrm{f}} = \sum_{\lambda=1}^{2} \sum_{\boldsymbol{k}} \hbar \boldsymbol{k} a^{\dagger}_{\boldsymbol{k},\lambda} a_{\boldsymbol{k},\lambda} \quad . \qquad (33)$$

The bound component is calculated in the same way as (19). However, in this case, the factor $\boldsymbol{y}$ is not present so the bound momentum is simply [8]

$$\boldsymbol{P}_{\mathrm{b}} = \frac{1}{c} \int \mathrm{d}^3 x \, \rho(\boldsymbol{x}, t) \boldsymbol{A}_{\mathrm{t}}(\boldsymbol{x}, t) \quad . \qquad (34)$$

The electromagnetic field variable $\boldsymbol{A}_{\mathrm{t}}$ may be quantized by using (24).

## 7. Conclusion

We have obtained a decomposition of the angular momentum of the classical electromagnetic field into orbital and spin components that is general and manifestly gauge invariant. This is done by decomposing the electric field into its longitudinal and transverse parts by means of the Helmholtz theorem. The orbital and spin components of the angular momentum of any specified electromagnetic field can be computed from this prescription. We find that the results agree with the known expressions for the angular momentum components of classical linearly and circularly polarized plane waves and plane waves quantized in the Coulomb gauge. It is not necessary to normal-order the Fock operators to obtain the quantum mechanical results.

## Appendix

The theorem of Helmholtz states that a 3-vector field may be decomposed into longitudinal and transverse parts

$$\boldsymbol{A}(\boldsymbol{r}) = \boldsymbol{A}_{\mathrm{l}}(\boldsymbol{r}) + \boldsymbol{A}_{\mathrm{t}}(\boldsymbol{r}) \qquad (\mathrm{A1})$$

with $\times \boldsymbol{A}_{\mathrm{l}}(\boldsymbol{r}) = \boldsymbol{0}$ and $.\boldsymbol{A}_{\mathrm{t}}(\boldsymbol{r}) = 0$ . (A2)

Let $\boldsymbol{A}_{\mathrm{l}}(\boldsymbol{r}) \rightarrow \boldsymbol{A}_{\mathrm{l}}'(\boldsymbol{r}) = \boldsymbol{A}_{\mathrm{l}}(\boldsymbol{r}) - \chi$ (A3a)

and $\boldsymbol{A}_{\mathrm{t}}(\boldsymbol{r}) \rightarrow \boldsymbol{A}_{\mathrm{t}}'(\boldsymbol{r}) = \boldsymbol{A}_{\mathrm{t}}(\boldsymbol{r}) + \chi$ . (A3b)

Hence $\times \boldsymbol{A}_{\mathrm{l}}'(\boldsymbol{r}) = \boldsymbol{0}$ and $.\boldsymbol{A}_{\mathrm{t}}'(\boldsymbol{r}) = {}^{2}\chi$ . (A4)

The decomposition will be maintained for solutions of ${}^{2}\chi = 0$. The solutions of ${}^{2}\chi = 0$ are of the forms $\chi = r^{l} \mathrm{Y}_{l}^{m}$ and $\chi = r^{-(l+1)} \mathrm{Y}_{l}^{m}$, where the *Y* are spherical harmonics and $0 \leq l, \quad -l \leq m \leq l$ [5]. The only solution that does not have a gradient that diverges at *r* = 0 or *r* = ∞ is the *l* = 1 term of the first set of solutions. This is of the form $\chi = \boldsymbol{r}.\boldsymbol{D}$ (***D*** uniform) and gives a uniform gradient ***D***. Therefore the Helmholtz decomposition is unique up to a uniform vector if divergent terms are not allowed.

Next consider the decomposition, with ***F*** being an arbitrary vector field,

$\boldsymbol{A}_{\mathrm{l}}(\boldsymbol{r}) \rightarrow \boldsymbol{A}_{\mathrm{l}}'(\boldsymbol{r}) = \boldsymbol{A}_{\mathrm{l}}(\boldsymbol{r}) + \times \boldsymbol{F}$ (A5a)

$\boldsymbol{A}_{\mathrm{t}}(\boldsymbol{r}) \rightarrow \boldsymbol{A}_{\mathrm{t}}'(\boldsymbol{r}) = \boldsymbol{A}_{\mathrm{t}}(\boldsymbol{r}) - \times \boldsymbol{F}$ (A5a)

Hence $.\boldsymbol{A}_{\mathrm{t}}'(\boldsymbol{r}) = 0$ and $\times \boldsymbol{A}_{\mathrm{l}}'(\boldsymbol{r}) = \times \times \boldsymbol{F}$ . (A6)

The decomposition continues to hold only if

$\times \times \boldsymbol{F} = 0$ . (A7)

A solution to (A7) is given by $\times \boldsymbol{F} = \phi$, where $\phi$ is a scalar field. By taking the divergence of both sides we obtain Laplace's equation for $\phi$. The arguments used above follow and again the Helmholtz decomposition remains unique up to a uniform vector.

## References

[1] Padgett, M.P. and Allen, L., 2000, Light with a twist in its tail, *Contemporary Physics,* **41**, 275-285.

[2] Allen, L., Beijersbergen, M.W., Spreeuw, R.J.C. and Woerdman, J.P., 1992, Orbital angular momentum of light and the transformation of Laguerre-Gaussian laser modes, *Physical Review A,* **45**, 8185-8189.

[3] Barnett, S.M., 2002, Optical angular momentum flux, *Journal of Optics B: Quantum and Semiclassical Optics,* **4**, S7-S16.

[4] Allen, L., Padget, M.J. and Babiker, M., 1999, in *Progress in Optics,* edited by Wolf, E. (Amsterdam: Elsevier) Vol. **39** p. 391-372.

[5] Jackson, J.D., 1999, *Classical Electrodynamics* 3rd edition (New York: Wiley).

[6] Ohanian, H.C., 1986, What is spin?, *American Journal of Physics,* **54**, 500-505.

[7] Gori, F., Santarsiero, M., Borghi, R. and Guatti, G., 1998, Orbital angular momentum of light: a simple view, *European Journal of Physics,* **19**, 439-444.

[8] Cohen-Tannoudji, C., Dupont-Roc, J. and Gilbert, G., 1989, *Photons and Atoms* (New York: Wiley).

[9] Arfken, G., 1995, *Mathematical Methods for Physicists* 4th edition (San Diego: Academic Press).

[10] Stewart, A.M., 2003, Vector potential of the Coulomb gauge, *European Journal of Physics,* **24**, 519-524.

[11] Stewart, A.M., 2004, Reply to Comments on "Vector potential of the Coulomb gauge", *European Journal of Physics,* **25**, L29-L30.

[12] Rohrlich, F., 2004, The validity of the Helmholtz theorem, *American Journal of Physics,* **72**, 412-413.

[13] Beth, R.A., 1936, Mechanical detection and measurement of the angular momentum of light, *Physical Review,* **50**, 115-125.

[14] Feynman, R.P., Leighton, R.B. and Sands, M., 1963, *The Feynman Lectures on Physics* (Reading MA.: Addison-Wesley). Ch 33, p10.

[15] Humblet, J., 1943, Sur le moment d'impulsion d'une onde electromagnetique, *Physica*, **10**, 585-603.

[16] Allen, L. and Padget, M.J., 2000, The Poynting vector in Laguerre-Gaussian beams and the interpretation of their angular momentum density, *Optics Communications*, **184**, 67-71.

[17] Khrapko, R.I., 2001, Question #79. Does plane wave not carry a spin?, *American Journal of Physics*, **69**, 405.

[18] Yurchenko, V.B., 2002, Answer to Question #79. Does plane wave not carry a spin?, *American Journal of Physics*, **70**, 568-569.

[19] Allen, L. and Padgett, M.J., 2002, Response to Question #79. Does plane wave not carry spin angular momentum?, *American Journal of Physics*, **70**, 567-568.

[20] Simmons, J.W. and Guttman, M.J., 1970, *States, Waves and Photons: A Modern Introduction to Light* (Reading, MA.: Addison-Wesley).

[21] Craig, D.P. and Thirunamachandran, T., 1998, *Molecular Quantum Electrodynamics: An Introduction to Radiation-Molecule Interactions* 2nt edition (New York: Dover).

[22] Greiner, W. and Reinhardt, J., 1996, *Field Quantization* (Berlin: Springer).

[23] Stewart, A.M., 2005, Orbital and spin components of the angular momentum of a general electromagnetic field. **Proceedings of the 16th National Biennial Congress of the Australian Institute of Physics,** Canberra, 31 Jan. - 4 Feb. (2005). ISBN 0-9598064-8-2. Paper AOS PWE 30.

[24] Stewart A.M., 2005, Angular momentum of the electromagnetic field: the plane wave paradox resolved. *European Journal of Physics*, **26** (4) 635-641 (2005).